\documentstyle[preprint, aps, epsfig,psfig]{revtex}
\tightenlines

\input epsf

\begin{document}
\draft
\title{Eliminating Thermal Violin Spikes from LIGO Noise}
\author{D.~H.~Santamore$^{(1),(2)}$ and Yuri Levin$^{(1),(3)}$}
\address{$^{(1)}$Theoretical Astrophysics and $^{(2)}$Condensed Matter Physics,\\
California Institute of Technology, Pasadena, CA 91125}
\address{$^{(3)}$Department of Astronomy, \\
601 Campbell Hall, University of California, Berkeley, CA 94720}
\date{printed \today}
\maketitle

\begin{abstract}
We have developed a scheme for reducing LIGO suspension thermal noise close
to violin-mode resonances. The idea is to monitor directly the
thermally-induced motion of a small portion of (a ``point'' on) each
suspension fiber, thereby recording the random forces driving the test-mass
motion close to each violin-mode frequency. One can then suppress the
thermal noise by optimally subtracting the recorded fiber motions from the
measured motion of the test mass, i.e., from the LIGO output. The proposed
method is a modification of an analogous but more technically difficult
scheme by Braginsky, Levin and Vyatchanin for reducing broad-band suspension
thermal noise.

The efficiency of our method is limited by the sensitivity of the sensor
used to monitor the fiber motion. If the sensor has no intrinsic noise (i.e.
has unlimited sensitivity), then our method allows, in principle, a complete
removal of violin spikes from the thermal-noise spectrum. We find that in
LIGO-II interferometers, in order to suppress violin spikes below the
shot-noise level, the intrinsic noise of the sensor must be less than $\sim
2\times 10^{-13}\hbox{cm}/\sqrt{\hbox{Hz}}$. This sensitivity is two orders
of magnitude greater than that of currently available sensors.
\end{abstract}


%\twocolumn
%*************************************************************

\section{introduction}

Suspension thermal noise is among the major sources of noise in the Laser
Interferometer Gravitational Wave Observatory\footnote{%
All results presented in this paper are equally applicable to LIGO's
international partners: VIRGO, GEO-600, TAMA, etc.} (LIGO) \cite{abramovici}%
. In mature interferometric gravitational wave detectors, such as those of
``LIGO-II'' interferometers, it is predicted to be significant in the broad
frequency band of 30-110Hz and to dominate other noise sources in the narrow
peaks around frequencies of the standing wave modes (so-called violin modes)
of the fibers on which the LIGO test masses are suspended. Fundamentally,
suspension thermal noise arises from randomly fluctuating stresses of
thermal origin in the suspension fibers. Currently, there are three general
strategies for reducing the suspension thermal noise:

1. Reduce the mechanical losses in the test-mass suspension, thereby
decoupling the suspension fiber's motions from the suspension's thermal
reservoir and thence reducing the fluctuating stresses in the suspension
fibers. There is a current vigorous experimental push in this direction
(see, e.g., \cite{braginsky}, \cite{rowan}, \cite{andri}).

2. Cool the suspension system \cite{Japangroup}, or some of its components%
\footnote{%
A combination of cooling of the top suspension point, and adjusting location
of the laser beam on the test mass to neutralize the thermal noise
originating at the bottom of the suspension fiber, may significantly reduce
suspension thermal noise in the broad frequency band \cite{BLV}.}. The above
two strategies will not be discussed in this paper.

3. Use displacement sensors to monitor the motion of the suspension fibers,
thereby recording independently the random forces responsible for the
Brownian motion of the test mass (Langevin forces), and then optimally
subtract this recorded force from the LIGO readout, thus achieving a partial
compensation of the thermal noise. The general idea of thermal noise
compensation has been widely discussed (See e.g.s Refs. \cite{kulagin} and 
\cite{weiss}). A concrete version of it for broad-band suspension thermal
noise in LIGO was conceived and analyzed by Braginsky, Levin, and Vyatchanin
(BLV) \cite{BLV}. Strigin and Vyatchanin have recently suggested a scheme
which in principle allows analogous compensation for internal thermal noise 
\cite{SV}.

A related approach to thermal noise compensation has been developed by
Heidmann et.~al.~\cite{heidemann} and Pinard et.~al.~\cite{pinard}. They
have demonstrated --- both experimentally and theoretically --- a partial
compensation of high-frequency internal thermal noise by implementing in
hardware the subtraction of the Langevin force from the readout through a
negative feedback loop. The result is an effective ``cooling'' of the
mirror's mechanical motions \footnote{%
It seems to us that Heidmann et al's \cite{heidemann} direct ``cooling''
technique requires a reference mirror which is much ``quieter'' than the
mirror used in LIGO, so it is not clear how practical such an experimental
arrangement is for LIGO. Nonetheless, \cite{heidemann} is an interesting
experimental demonstration of the principle of dynamical thermal noise
compensation.}.

In BLV, the broad-band 30-110 Hz component was targeted for reduction using
a scheme where the motion of a suspension fiber is monitored along all its
length, and an optimized average displacement is recorded and subtracted
from the LIGO data. This scheme would be difficult to implement in practice,
due to the fraction-of-a-micron proximity of the optical waveguide-based
sensor and the fiber. In this paper, we instead propose to track the fiber's
motion at a single point, which should be significantly easier to implement.
This single-point-tracking does not allow significant reduction of the
broad-band component of the suspension thermal noise, but when combined with
optimal subtraction it should be quite effective in removing violin peaks
from the thermal noise spectrum. The possibility of compensation of the
thermal noise at violin resonances was briefly pointed out by Pinard et al.~ 
\cite{pinard} in the context of the ``cooling'' technique.

Distinct from {\it reduction} of the suspension thermal noise, there are
viable strategies to {\it filter out} narrow thermal violin-mode noise from
the LIGO data \cite{Finn}. By contrast with our scheme, a procedure of this
kind would filter out {\it both} the noise and the signal around the violin
spikes. This works well in most cases, since violin modes dominate other
noise in very narrow frequency bands, and it is not likely that the signal
in these particular bands would be significant. However, if the shot noise
is reduced by increasing the laser power or by narrow-banding the
interferometer's response, the broad bases of violin spikes could become
dominant, and the genuine thermal noise compensation scheme, such as
proposed here, could well be of use.

This paper is structured as follows: in Sec. II we give an intuitive
introduction to the origin of suspension thermal noise and explain
qualitatively why our method of thermal noise compensation should work.

In Sec. III we formally introduce a new readout variable, $x_{{\rm readout}%
}=x_{{\rm testmass}}+\alpha x_{{\rm fiber}}$, where $x_{{\rm testmass}}$ and 
$x_{{\rm fiber}}$ are displacements of the test mass and of the
independently monitored point of the fiber, and $\alpha $ is a
frequency-dependent number which must be chosen in an optimal way. We then
show how to use the Fluctuation-Dissipation theorem to compute the thermal
noise in this new readout variable.

Section IV presents results of our calculations of the optimized parameter $%
\alpha $ and of the reduced thermal noise. We find (see Fig.\ \ref{graph})
that on resonance the optimal value of $\alpha $ is $(1/\pi )m/M$, where $m$
and $M$ are the masses of the suspension fiber and the test mass,
respectively.

In Sec. V we deal with the practical issue of the imperfection of the sensor
which is used to monitor the suspension fiber. We find that if the sensor is
infinitely precise (i.e., if it introduces no noise to the measurement),
then the violin spike could be removed completely from the spectrum;
however, in most realistic situations the effectiveness of our method will
be severely limited by the sensor noise. For example, in order to bring down
a violin spike to the shot-noise level at the violin frequency, one needs a
sensor with the sensitivity $\sim 1.8\times 10^{-13}{\rm cm}/\sqrt{\hbox{Hz}}
$ [cf.~Eq.~(\ref{nsf2}) of the text]; this sensitivity is significantly
better than $\sim 10^{-11}{\rm cm}$, which is the best sensitivity of
currently used interferometric sensors of fiber motion \cite{ageev}. Perhaps
the situation will change when the next generation of displacement sensors
comes to life.

%*********************************************************************

\section{Motivation and intuition}

In this section we discuss intuitively the origin of suspension thermal
noise and how to compensate it by monitoring devices.

Suspension thermal noise can be traced to randomly fluctuating stresses
which are located in each suspension fiber. These stresses make the
suspension fiber move in a random way; the fiber will make its test mass
move randomly as well by pulling it sideways. It is this resulting random
motion of the test mass that is referred to as the ``suspension thermal
noise''.

Clearly, the motion of the fiber and the test mass are not independent. The
fiber's random displacement is $\sim M/m$ greater than that of the test
mass, where $M$ and $m$ are the masses of the test mass and the fiber
respectively. BLV have argued that by monitoring the random motion of the
fiber one gains information about the random forces acting on the test mass;
one can then use this information to effectively reduce suspension thermal
noise by orders of magnitude, at least in principle. Their scheme requires
to measure $x_{{\rm fiber}}$, which is a horizontal displacement
approximately averaged over the fiber, then construct the new readout
variable 
\begin{equation}
x_{{\rm readout}}=x_{{\rm testmass}}+\alpha x_{{\rm fiber}},
\label{xreadout}
\end{equation}
where $\alpha $ is an optimization factor of order $m/M$ that must be
calculated theoretically. It is this new readout variable which has low
thermal noise: The gravitational-wave signal is virtually unchanged by using
this new readout variable, since $\alpha $ is very small ($8\times 10^{-6}$
for LIGO).

The BLV scheme facilitates reduction of suspension thermal noise in the
broad-band range of frequencies between the pendulum and the first violin
peak, and if needed between the violin peaks. At the same time, their scheme
has the disadvantage of requiring precise monitoring of the motion of the
suspension fiber along all of its length\footnote{%
Of course, one must monitor all of the suspension fibers.}; at the moment
there is, to our knowledge, no active experimental effort in this direction.

In this paper we propose a modification of the BLV scheme. The new scheme is
easier to implement experimentally because it requires monitoring the fiber
at a single point somewhere near the middle, rather than over all of its
length. The requirement for the sensor sensitivity is also significantly
less stringent than in the BLV proposal. However, single-point monitoring is
not effective at removing the thermal noise in the broad-band region.
Instead, our scheme is designed only to remove the thermal noise around the
``violin spikes'', which pierce the photon shot-noise floor at frequencies
above $\sim 300$Hz for LIGO.

If we restrict our vision to a narrow band of frequencies close to some
particular violin resonance, we will find that the motion of the fiber is
remarkably simple: it looks like a standing wave with an integer number of
half-wavelengths fitting the length of the fiber. The amplitude of the wave
is a thermally fluctuating variable, but the shape does not fluctuate much.
By sensing the fiber motion at a single point, we can monitor the
fluctuating amplitude of the violin-resonance standing wave, and thus infer
the random force with which the fiber pulls sideways on the test mass, at a
frequency close to the violin resonance.

As an example, consider the dynamics of a suspension fiber and its test mass
close to the first violin resonance, which corresponds to a violin mode with
half-wavelength equal to the length of the fiber. The fiber's horizontal
displacement is given by 
\begin{equation}
x_{{\rm f}}(z)=A\sin (kz)\sin (\omega _{{\rm v}}t),  \label{xf}
\end{equation}
where $A$ is the (random) amplitude of the standing wave, $z$ is distance
from a point on the fiber to the top of the fiber, $k$ is the wave vector,
such that $kl=\pi $ (with $l$ the length of the fiber), and $\omega _{{\rm v}%
}=(\pi /l)\sqrt{Mgl/m}$ is the angular frequency of the first violin
resonance. The force with which the fiber pulls sideways on the test mass is
given by 
\begin{equation}
F_{{\rm f}}=-Mg\left( {\frac{dx_{{\rm f}}}{dz}}\right) _{z=l}=Mg\pi {\frac{A%
}{l}}\sin (\omega _{{\rm v}}t).
\end{equation}

Assume for simplicity that the test mass has no tilt degree of freedom. This
is actually the case for the currently planned 4-fibers suspension on each
LIGO-II test mass. Then the laser beam spot measuring the test-mass position
is displaced by the same amount as the test-mass center of mass: 
\begin{equation}
x_{{\rm testmass}}=-{\frac{F_{{\rm f}}}{M\omega _{{\rm v}}^{2}}}=-{\frac{m}{%
\pi M}}A\sin (\omega _{{\rm v}}t).  \label{xtestmass}
\end{equation}
We neglect the test mass radius compared to the wire length. Equation (\ref
{xtestmass}) relates the thermal motion of the fiber, which is almost
completely determined by fluctuations in $A$, to the thermal motion of the
test mass, at a frequency close to $\omega _{{\rm v}}$. If we choose to
monitor the fiber at its midpoint, i.e., $x_{{\rm fiber}}=x_{{\rm f}}(z=l/2)$%
, then the linear combination 
\begin{equation}
x_{{\rm readout}}=x_{{\rm testmass}}+{\frac{m}{\pi M}}x_{{\rm fiber}}
\label{xreadout1}
\end{equation}
is expected to have significantly reduced thermal noise at the first violin
peak; cf.\ Eq.\ (\ref{xreadout}).

How much can the thermal noise be reduced when this new readout is
introduced? What is the optimal readout variable for frequencies slightly
off the violin resonance? In the next subsection we use the
Fluctuation-Dissipation theorem to answer these questions.

\section{The new readout variable: issues of principle}

For simplicity, let us assume that the test mass is hanging on a single
fiber, and has no tilt degrees of freedom; see Fig.~\ref{generalizedforce}.
In LIGO-II, the test mass is suspended on two fiber loops, which from the
point of view of suspension thermal noise is equivalent to the test mass
suspended on four independent fibers, and in this configuration the no-tilt
assumption is realistic. The generalization of our scheme and analysis to a
multi-fiber suspension is straightforward.

We assume that the gravitational-wave interferometer monitors the
displacement of the test mass \footnote{%
Here it is assumed that the test mass moves as a rigid body, that is, we
neglect the internal thermal noise due to fluctuations of the test-mass
shape.} $x_{{\rm testmass}}$, and that an independent sensor $S$ measures
the displacement of the suspension fiber $x_{{\rm fiber}}$ at some point $%
z_{0}$.

The observer should then record a new readout variable, 
\begin{equation}
x_{{\rm readout}}=x_{{\rm testmass}}+\alpha x_{{\rm fiber}},
\label{xreadout2}
\end{equation}
where $\alpha$ is a frequency-dependent coefficient which has to be chosen
so that the thermal noise in $x_{{\rm readout}}$ is minimized.

The Fluctuation-Dissipation theorem \cite{CW} allows one to calculate the
spectral density $S_{x}(f,\alpha )$ of the thermal noise in $x_{{\rm readout}%
}$, where $f$ is the frequency; see Refs.\cite{Yuri} and \cite{BLV} for the
general method of calculation, and see \cite{gonzalez} for the first direct
application of the Fluctuation-Dissipation theorem to computation of
suspension thermal noise. First, we imagine that a generalized oscillating
force $F(t)=F_{0}\cos (\omega t)$ conjugate to $x_{{\rm readout}}$ is
applied to the fiber $+$ test mass system (here and elsewhere in this paper $%
\omega =2\pi f$). We introduce such a force via the interaction Hamiltonian, 
\begin{equation}
H_{{\rm int}}=-F(t)x_{{\rm readout}}.  \label{Hint}
\end{equation}
As discussed in \cite{BLV}, and as is apparent from Eqs. (\ref{xreadout2})
and (\ref{Hint}), applying the generalized force $F$ is equivalent to
applying two simple ``Newtonian'' forces simultaneously: one, with a
magnitude $F(t)$, to the test mass, and the other, with a magnitude $\alpha
F(t)$, at a point $z=z_{0}$ on the fiber.

The next step is to calculate the motion of the test mass and the suspension
fiber under the action of the generalized force, $F(t)$, and to work out the
average power, $W_{{\rm diss}}$, that would be dissipated as heat as a
result of such motion. The thermal noise in $x_{{\rm readout}}$ is then
given by 
\begin{equation}
S_{x}(f,\alpha )={\frac{8k_{B}T}{\omega ^{2}}}{\frac{W_{{\rm diss}}}{%
F_{0}^{2}}},  \label{Sxfalpha}
\end{equation}
where $k_{B}$ is the Boltzmann constant, and $T$ is the temperature of the
suspension fiber (cf. Eq.\ (3.10) of \cite{Yuri}).

The last step is to choose $\alpha$ such that $S_x(f,\alpha)$ is as small as
possible.

The oscillatory motion of the fiber$+$test mass under the action of the
generalized force $F$ is shown in Fig.\ \ref{generalizedforce}. The fiber is
strongly bent near three points\footnote{%
The length over which the fiber is bent is given by $\lambda =\sqrt{JE/Mg}$,
where $J$ is the geometric moment of inertia for the fiber crossection (it
equals to $\pi d^{4}/64$ for the circular crossection of diameter $d$), and $%
E$ is the Young modulus of the fiber material. For typical LIGO parameters $%
\lambda $ is a small fraction of a centimeter, much lass than the length of
the suspension fiber}: at the bottom and top attachment points, and at the
point on the fiber where we imagine applying $\alpha F$ (the location of the
independent sensor). It is near these three points the regions in which the
fiber bends and heat is produced by the fiber$+$test mass motion.

In order to compute the dissipated power $W_{{\rm diss}}$, and then to use
it in Eq.~(\ref{Sxfalpha}) to calculate the thermal noise, one must specify
a model for dissipative losses in the suspension fiber. We assume that the
time-averaged power dissipated as heat is given by 
\begin{equation}
W_{{\rm diss}}=f\left[ \zeta _{{\rm top}}\theta _{{\rm top}}^{2}+\zeta _{%
{\rm bottom}}\theta _{{\rm bottom}}^{2}+\zeta _{{\rm middle}}\theta _{{\rm %
middle}}^{2}\right] ,  \label{wdiss}
\end{equation}
where $\theta _{{\rm top}}$, $\theta _{{\rm bottom}}$ and $\theta _{{\rm %
middle}}$ are the amplitudes of the oscillating top, bottom and middle
angles respectively, see Fig.\ \ref{generalizedforce}, and $\zeta _{{\rm top}%
}$, $\zeta _{{\rm bottom}}$ and $\zeta _{{\rm middle}}$ are
frequency-dependent constants which are determined by the dissipation
mechanism. If the source of dissipation is distributed homogeneously, then

\begin{equation}
\zeta_{{\rm top}}=\zeta_{{\rm bottom}}=2\zeta_{{\rm middle}}=\zeta.
\label{zeta}
\end{equation}

The amplitudes of the oscillating angles $\theta _{{\rm top}}$, $\theta _{%
{\rm bottom}}$ and $\theta _{{\rm middle}}$ are worked out in Appendix A:

\begin{equation}
\theta _{top}={\frac{F_{0}}{{M}}}\left[ {\frac{k\left[ 1+\alpha \cos \left[
k\left( l-z_{0}\right) \right] \right] -\left( {\alpha /g}\right) {\omega }%
^{2}\sin \left[ {k}\left( {l-z_{0}}\right) \right] }{{gk\cos }\left( {kl}%
\right) {-\omega ^{2}\sin }\left( {kl}\right) }}\right] ,  \label{thetatop}
\end{equation}

\begin{equation}
\theta _{bottom}={\frac{F_{0}}{{M}}}\left[ {\frac{{k\cos }\left( {kl}\right)
+\left( \alpha /g\right) \omega ^{2}\sin \left( kz_{0}\right) }{{kg\cos }%
\left( {kl}\right) {-\omega ^{2}\sin }\left( {kl}\right) }}\right] ,
\label{thetabottom}
\end{equation}
and 
\begin{equation}
\theta _{middle}={\frac{\alpha F_{0}}{Mg}}.  \label{thetamiddle}
\end{equation}
Here $M$ is the mass of the mirror, $m$ is the mass of the suspension fiber, 
$l$ is the fiber length, $z_{0}$ is the distance along the fiber from the
top attachment point to the sensor, 
\begin{equation}
k=\omega /\sqrt{Mgl/m},  \label{wavenumber}
\end{equation}
is the wavenumber of the standing wave induced in the fiber, and $z_{0}$ is
the distance from the suspension top to the sensor.

The expressions for $\theta _{{\rm top}}$ and $\theta _{{\rm bottom}}$ do
not take into account damping, and hence they diverge when $\omega $
approaches the violin-resonance angular frequency. We correct for this by
replacing $\omega $ and $k$ by $\omega (1+i/Q_{n})$ and $k(1+i/Q_{n})$ when $%
\omega $ is close to the $n$-th violin-resonance\footnote{%
We can justify this procedure by decomposing the motion of suspension fiber
into normal modes, and considering how each mode is driven by the
generalized force. When the driving frequency is close to a proper frequency
of some violin mode, we can neglect excitation of other modes. Complexifying
the frequency in the way described above is the standard procedure for
finding the response of a damped harmonic oscillator close to its resonance
frequency. Here by ``close'' we mean within a few widths $\gamma _{n}=\omega
_{n}/Q_{n}$ of the $n$-th violin resonance; in our numerical evaluations we
used the complex replacement within $10\gamma _{n}$ from $n$-th resonance.},
and by taking the absolute value of the now complex expressions for $\theta
_{{\rm top}}$ and $\theta _{{\rm bottom}}$. Here $Q_{n}$ is the quality
factor of the $n$-th violin mode; it can be shown, using Eqs.~(\ref{wdiss})
and (\ref{zeta}), that 
\begin{equation}
Q_{n}={\frac{\pi Mgl}{2\zeta (f_{n})}},  \label{Qn}
\end{equation}
for the case when the damping is homogeneously distributed in the fiber and
is characterized by a single parameter $\zeta(f) $.

We can then use Eqs. (\ref{thetatop}), (\ref{thetabottom}) and (\ref
{thetamiddle}) to compute the dissipated power, $W_{{\rm diss}}$, from Eq.~(%
\ref{wdiss}). To minimize the readout thermal noise $S_{x}(f,\alpha )$, we
must choose the optimal $\alpha =\alpha _{{\rm opt}}$ which minimizes $W_{%
{\rm diss}}$, i.e. $\partial W_{{\rm diss}}/\partial \alpha =0$. 
%************************************************************

\section{The new readout variable: Results}

When we carry through the calculations outlined in the previous section, we
find that the optimal readout variable which minimizes the thermal noise is 
\begin{equation}
x_{{\rm opt}}=x_{{\rm mirror}}+\alpha _{{\rm opt}}x_{{\rm fiber}},
\label{xoptimal}
\end{equation}
where 
\begin{eqnarray}
\alpha _{{\rm opt}} &=&A_{11}/\left( B_{11}+B_{12}+B_{13}\right) ;  \nonumber
\\
A_{11} &=&-2gk\left\{ gk\cos \left[ k\left( l-z_{0}\right) \right] +\omega
^{2}\left[ \cos \left( kl\right) \sin \left( kz_{0}\right) -\sin \left[
k\left( l-z_{0}\right) \right] \right] \right\}  \label{alphaopt} \\
B_{11} &=&g^{2}k^{2}\cos ^{2}\left( kl\right) +2g^{2}k^{2}\cos ^{2}\left[
k\left( l-z_{0}\right) \right]  \nonumber \\
B_{12} &=&\omega ^{2}\left[ \omega ^{2}\sin ^{2}\left( kl\right) -gk\sin
\left( 2kl\right) +2\omega ^{2}\sin ^{2}\left[ k\left( l-z_{0}\right) \right]
\right] \\
B_{13} &=&2\omega ^{2}\left[ \omega ^{2}\sin ^{2}(kz_{0})-gk\sin
[2k(l-z_{0})]\right] .  \nonumber
\end{eqnarray}
The optimal value $\alpha _{{\rm opt}}$ is plotted in Fig.\ \ref
{optimizedalpha} for the case when the sensor is monitoring the midpoint of
the fiber, i.e. $z_{0}=l/2$. As is clear from the figure, $\alpha _{{\rm opt}%
}$ is frequency-dependent, and one should keep this in mind while designing
the numerical procedure for the data analysis. On resonance (where our
method is expected to be most efficient), we get

\begin{equation}
\alpha _{{\rm opt}}\simeq\frac{m}{\pi M}=6.36\times 10^{-6} \left({\frac{m}{%
0.3\hbox{g}}}\right)\left({\frac{15\hbox{kg}}{M}}\right),
\end{equation}
which is in exact agreement with our more intuitive calculation in Eq.~(\ref
{xreadout1}).

The thermal noise with and without the optimal monitoring, $S_{x}(f,\alpha _{%
{\rm opt}})$ and $S_{x}(f,\alpha =0)$ are plotted in Fig.\ \ref{graph}
(solid and dashed lines respectively). We see that the optimal monitoring
removes completely the first and other ``odd'' violin spikes from the
suspension thermal noise spectrum. ``Even'' violin spikes can also be
removed if the sensor is positioned off-center of the suspension fiber.

How precisely do we need to choose $\alpha $, so that significant noise
reduction is achieved? The spectral density $S_x(f,\alpha)$ is a quadratic
function of $\alpha$. We can use the fact that it has a minimum at $%
\alpha=\alpha_{{\rm opt}}$ to write 
\begin{equation}
S_{x}(f,\alpha )=S_{x}(f,\alpha _{{\rm opt}})+{\Delta S(f)}\left( {\frac{%
\alpha -\alpha _{{\rm opt}}}{\alpha _{{\rm opt}}}}\right) ^{2},
\label{sxalpha1}
\end{equation}
where $\Delta S(f)=S_{x}(f,\alpha =0)-S_{x}(f,\alpha _{{\rm opt}})$. When
the noise reduction is effective, $\Delta S(f)\simeq S_{x}(f,\alpha =0)$.
From Eq.~(\ref{sxalpha1}) we see that so long as 
\begin{equation}
\left| {\frac{\alpha -\alpha _{{\rm opt}}}{\alpha _{{\rm opt}}}}\right| <%
\sqrt{\frac{S_{x}(f,\alpha _{{\rm opt}})}{S_{x}(f,\alpha =0)}},
\label{criterialpha}
\end{equation}
the thermal noise reduction will not be seriously compromised. If the above
condition is not satisfied, then the thermal noise is reduced by a factor $%
\sim (\alpha -\alpha _{{\rm opt}})^{2}/\alpha _{{\rm opt}}^{2}$ which
depends only on how well the observer is able to tune $\alpha $.

\section{Influence of the sensor noise}

So far we have assumed that the sensor, which is independently monitoring
the fiber displacement, has an infinite precision, i.e.\ we have assumed
that it does not introduce any noise of its own into the readout variable.
In real life no sensor is perfect. The sensor readout is, in general, given
by 
\begin{equation}
x_{{\rm sensor}}=x_{{\rm fiber}}+n_{s},  \label{xsensor}
\end{equation}
where $n_{s}$ is the random noise introduced by the sensor. We assume that $%
n_{s}$ is a Gaussian random variable with noise spectral density, $N_{s}(f)$.

The total readout variable is then 
\begin{equation}
x_{{\rm readout}}=x_{{\rm mirror}}+\alpha (x_{{\rm fiber}}+n_{s}),
\label{xreadout3}
\end{equation}
and the total readout noise is then 
\begin{equation}
S_{{\rm total}}(f,\alpha )=S_{x}(f,\alpha )+\alpha ^{2}N_{s}(f).
\label{totalnoise1}
\end{equation}
A new optimal value $\alpha _{{\rm new}}$ of $\alpha $ has to be evaluated,
so that $\partial S_{{\rm total}}(f,\alpha )/\partial \alpha =0$. Using
Eqs.~(\ref{sxalpha1}) and (\ref{totalnoise1}), we get after straightforward
algebra: 
\begin{equation}
\alpha _{{\rm new}}(f)={\frac{\Delta S(f)}{\Delta S(f)+\alpha _{{\rm opt}%
}^{2}(f)N_{s}(f)}}\alpha _{{\rm opt}}(f).  \label{alphanew}
\end{equation}
The optimized noise in the total readout variable is then 
\begin{equation}
S_{{\rm total}}(f,\alpha _{{\rm new}})=S_{x}(f,\alpha _{{\rm opt}})+%
\hbox{Harm}\{\Delta S(f),N_{s}(f)\alpha _{{\rm opt}}^{2}(f)\},  \label{snew1}
\end{equation}
where $\hbox{Harm}\{A,B\}=AB/(A+B)$ is a harmonic mean of $A$ and $B$. We
use the above equation to compute the solid curve in Fig.\ \ref{graph}.

In the neighborhood of the violin resonance, $\Delta S(f)\simeq
S_{x}(f,\alpha =0)$. The criterion for achieving significant noise reduction
is then 
\begin{equation}
N_{s}(f)\ll S_{x}(f,\alpha =0)/\alpha _{{\rm opt}}^{2}(f).  \label{nsf}
\end{equation}
On resonance \cite{saulson}, 
\begin{equation}
S_{x}(f_{v},\alpha =0)={\frac{4k_{B}TQm}{M^{2}\omega _{v}^{3}n^{2}\pi ^{2}}}
\end{equation}
and $\alpha _{{\rm opt}}=(1/\pi )m/M$ (here $n$ is the order of the violin
mode). Then a first requirement for the sensor sensitivity\footnote{%
Essentially, this requirement amounts to what one would expect intuitively:
the sensor must be able to resolve the thermal motion of the fiber.} (that
is, the sensitivity for which significant noise reduction can be achieved on
resonance) in Eq.~(\ref{nsf}) becomes 
\begin{equation}
\sqrt{N_{s}(f_{v})}\ll {\frac{1}{n\pi }}\sqrt{\frac{k_{B}TQ}{2\pi mf_{v}^{3}}%
}\simeq 4.2\times 10^{-7}{\hbox{cm}/\sqrt{\hbox{Hz}}}\left( {\frac{0.3%
\hbox{g}}{m}}\right) ^{1/2}\left( {\frac{Q}{10^{6}}}\right) ^{1/2}\left( {%
\frac{500\hbox{Hz}}{f_{v}}}\right) ^{3/2}\left( {\frac{T}{300\hbox{K}}}%
\right) ^{1/2}.  \label{nsf1}
\end{equation}

Most sensors are broad-band, and their sensitivity is essentially $f$%
-independent for the range of detectable gravitational-wave frequencies. The
precision of monitoring specified by Eq.~(\ref{nsf1}) is achievable by
currently available shadow sensors, which can measure with precision as high
as $10^{-9}\hbox{cm}/\sqrt{\hbox{Hz}}$; therefore violin spikes can be
reduced somewhat even using current technology.

However, we would ideally like to reduce the thermal violin spike below the
shot noise level. Shot noise will dominate over all other noise sources
except its violin spikes, at the violin-spike frequencies. When one
represents it by a random displacement of an individual test mass, this shot
noise for LIGO-II interferometers is \cite{LIGOspec} 
\begin{equation}
\sqrt{S_{{\rm shotnoise}}(f)}\simeq 1\times 10^{-18}\hbox{cm}/\sqrt{\hbox{Hz}%
}\times \left( {\frac{f}{500\hbox{Hz}}}\right) .  \label{shotnoise}
\end{equation}
The sensor sensitivity requirement in Eq.~(\ref{nsf}) then becomes 
\begin{equation}
\sqrt{N_{s}(f_{v})}<\sqrt{S_{{\rm shotnoise}}(f_{v})}/\alpha _{{\rm opt}%
}\simeq 2\times 10^{-13} \hbox{cm}/\sqrt{\hbox{Hz}}.  \label{nsf2}
\end{equation}
Ageev, Bilenko, and Braginsky \cite{ageev} have recently built an
interferometer which can measure a fiber displacement with precision of $%
\sim 10^{-11}\hbox{cm}/\sqrt{\hbox{Hz}}$. They are hoping to improve this
sensitivity by two orders of magnitude in the near future \cite{bilenko}.
This would, in principle, make an interferometric sensor viable for our
scheme.

However, in a narrow-band search the sensor-noise requirement is more severe
than Eq.~(\ref{nsf2}) by a factor $\sim 5$, because the shot-noise spectral
density is lower.

\section{discussion and Conclusions}

Among possible sources of periodic gravitational waves, the most promising
are neutron stars with spin periods of $1-5$ milliseconds \cite{millisecond}%
\footnote{%
This statement is based on current astronomical observations which use
electromagnetic waves (light, radio waves, x-rays, etc.). Naturally, LIGO
might change our notion about what the most promising gravitational-wave
sources are.}. The frequencies of these stars' gravitational waves lie in
the range which is pierced by the thermal violin spikes. Detection of these
gravitational waves by LIGO will require changing the reflectivity and the
location of the signal recycling mirror of the interferometer in such a way
that the shot noise is reduced significantly in a narrow band around the
gravitational-wave frequency. If a violin spike happens to be nearby,
eliminating it might help the gravitational-wave detection and measurement.
Our method allows one to completely eliminate violin spikes from the thermal
noise spectrum, at least in principle. Figure.\ \ref{graph} illustrates this
by showing a suspension thermal noise curve with and without compensation.

In practice, though, we seek a non-trivial experimental design in which each
suspension fiber is monitored at a single point by an independent sensor.
High sensor sensitivity is the key for the method to be effective. For
LIGO-II we need the noise spectral density introduced by each sensor to be
less than $\sim 2\times 10^{-13}\hbox{cm}/\sqrt{\hbox{Hz}}$; this would
allow us to reduce the violin spike below the broad-band shot-noise
sensitivity curve. In order to ``bring'' the violin spike below the {\it %
narrow-band} shot-noise sensitivity curve, we need even higher sensor
sensitivity, by a factor of $\sim 5$. Two types of sensors are currently
being used for detection of the fiber displacement: the shadow sensors and
the interferometric displacement sensors. The shadow sensors of $\sim 10^{-9}%
\hbox{cm}/\sqrt{\hbox{Hz}}$ have been demonstrated, and $10^{-10}\hbox{cm}/%
\sqrt{\hbox{Hz}}$ should be possible \cite{david}; but this still is a far
call from what we need for effective thermal noise compensation.
Interferometric sensors, such as those developed at Moscow State University,
look more promising. Ageev, Bilenko, and Braginsky \cite{ageev} have
achieved a sensitivity of $\sim 10^{-11}\hbox{cm}/\sqrt{\hbox{Hz}}$ for a
sensor based on reflecting light from a polished spot on a steel wire. The
current experiment in Moscow \cite{bilenko} hopes to achieve significantly
higher sensitivity by attaching a tiny mirror at the center of the fiber%
\footnote{%
The mirror might introduce extra mass and mechanical friction to the center
of the fiber, which would increase thermal noise in our readout variable.
Techniques developed in this paper can be readily used to analyse
theoretically thermal noise for such experimental set-up.}. It remains to be
seen whether future interferometric sensor designs will allow our method to
be practical for LIGO.

\section{acknowledgement}

We thank Kip Thorne for helpful advice, and for comments on this manuscript.
DHS gratefully acknowledges Michael Cross for useful discussions and support
while this paper was being written. Vladimir Braginsky and David Shoemaker
have provided us with useful information about interferometric and shadow
sensors, respectively. YL thanks ITP at UC Santa Barbara for hospitality
during his extended visit. This work is supported by NSF Physics grant,
PHY-9900997, YL also has been supported by the Theoretical Astrophysics
Center at UC Berkeley, and by NSF physics grant PHY-9907949 at ITP. 
%****************************************************

\appendix

\section{motion of the periodically driven suspension fiber}

In this appendix we derive the expressions for $\theta _{{\rm top}}$, $%
\theta _{{\rm bottom}}$ and $\theta _{{\rm middle}}$, Eqs. (\ref{thetatop}),
(\ref{thetabottom}), and (\ref{thetamiddle}) of the main text. These are the
amplitudes of the oscillating top, bottom and middle bending angles
respectively (see Fig.\ \ref{generalizedforce}), when a periodic force $%
F=F_{0}\cos (\omega t)$ is applied to the mirror, and simultaneously a force 
$\alpha F$ is applied to the point $z=z_{0}$ on the fiber at which the
displacement sensor makes its measurement. In our derivation we follow the
spirit of Appendix A in BLV.

For convenience, we complexify the driving force: 
\begin{equation}
F=F_{0}e^{i\omega t}.  \label{fcomplex}
\end{equation}
The equation of motion of the fiber with a force $\alpha F_{0}e^{i\omega t}$
applied at $z=z_{0}$ is

\begin{equation}
{\frac{\partial ^{2}{x_{{\rm f}}}}{\partial {t}^{2}}}=c^{2}{\frac{\partial
^{2}{x_{{\rm f}}}}{\partial {z}^{2}}}+{\frac{\alpha F_{0}}{{\rho }}}\delta
(z-z_{0})e^{i\omega t},  \label{eq:wave}
\end{equation}
where $x_{{\rm f}}(z,t)$ is the horizontal displacement of the fiber at
point $z$ and time $t$, $c$ is the velocity of a wave in the fiber, $c=\sqrt{%
glM/m}$, $M$ is the mass of the test mass, $m$ is the mass of the fiber, $l$
is the length of the fiber, $g$ is the earth's gravity, and $\rho $ is the
fiber's mass per unit length.

We look for a solution of Eq.~(\ref{eq:wave}) in the form $x_{{\rm f}%
}(z,t)=u(z)e^{i\omega t}$. Then for $0<z<z_{0}$ we have 
\begin{equation}
u(z)=A\sin \left( {kz}\right) ,  \label{A}
\end{equation}
and for $z_{0}<z<l$ we have 
\begin{equation}
u(z)=B\sin \left( {kz}\right) +C\cos \left( {kz}\right) ,  \label{B}
\end{equation}
where $A$, $B$, and $C$ are constants to be determined, and $k=\omega /c$ is
the wavenumber of the standing wave induced in the fiber.. The jump
conditions at $z=z_{0}$ are:

\begin{eqnarray}
&&u(z_{0}+\epsilon )=u(z_{0}-\epsilon ), \\
&&\left. {\frac{du}{{dy}}}\right| _{z_{0}+\epsilon }-\left. {\frac{du}{{dy}}}%
\right| _{z_{0}-\epsilon }={\frac{\alpha F_{0}}{{gM}}.}
\end{eqnarray}
Applying these jump conditions to Eqs (\ref{A}) and (\ref{B}), we get 
\begin{equation}
A=B+C\cot (kz_{0}),C={\frac{\alpha F_{0}}{Mgk}}\sin (kz_{0}).  \label{A1}
\end{equation}
To close this system of equations, we use Newton's second law for the test
mass, projected onto the horizontal axis [cf. Eq.~(35) of BLV]: 
\begin{equation}
-\omega ^{2}u(l)+g\left( {\frac{\partial u(z)}{\partial z}}\right) _{z=l}={%
\frac{F_{0}}{M}}.  \label{secondlaw}
\end{equation}
By solving together Eqs.~(\ref{A1}) and (\ref{secondlaw}), we get 
\begin{equation}
A={\frac{F_{0}}{M}}\left[ {\frac{1+\alpha \cos \left[ k\left( l-z_{0}\right) %
\right] -\left( {\alpha /gk}\right) \omega ^{2}\sin \left[ k\left(
l-z_{0}\right) \right] }{gk\cos \left( kl\right) -\omega ^{2}\sin \left(
kl\right) }}\right] ,  \label{B11}
\end{equation}
\begin{equation}
B={\frac{F_{0}}{M}}{\frac{1+\alpha \lbrack \sin (kl)\sin (kz_{0})+(\omega
^{2}/gk)\sin (kz_{0})\cos (kl)]}{gk\cos (kl)-\omega ^{2}\sin (kl)}},
\label{B1}
\end{equation}
\begin{equation}
C={\frac{F_{0}\alpha }{Mgk}}\sin (kz_{0})  \label{C!}
\end{equation}
Using Eqs.\ (\ref{A}), (\ref{B}), and (\ref{B11})-(\ref{C!}), we can now
work out the fiber's three angles of bent; 
\begin{equation}
\theta _{top}=\left. {{\frac{\partial {x}}{{\partial {z}}}}}\right| _{z=o}={%
\frac{F_{0}}{{M}}}\left[ {\frac{k\left[ 1+\alpha \cos \left[ k\left(
l-z_{0}\right) \right] \right] -\left( {\alpha /g}\right) {\omega }^{2}\sin %
\left[ {k}\left( {l-z_{0}}\right) \right] }{{gk\cos }\left( {kl}\right) {%
-\omega ^{2}\sin }\left( {kl}\right) }}\right] ,  \label{thetatop1}
\end{equation}

\begin{equation}
\theta _{bottom}=\left. {{\frac{\partial {x}}{{\partial {z}}}}}\right|
_{z=l}={\frac{F_{0}}{{M}}}\left[ {\frac{{k\cos }\left( {kl}\right) +\left(
\alpha /g\right) \omega ^{2}\sin \left( kz_{0}\right) }{{kg\cos }\left( {kl}%
\right) {-\omega ^{2}\sin }\left( {kl}\right) }}\right] ,
\label{thetabottom1}
\end{equation}
and 
\begin{equation}
\theta _{middle}=\left. {{\frac{\partial {x}}{{\partial {z}}}}}\right|
_{z=z_{0}-\epsilon }{-}\left. {{\frac{\partial {x}}{{\partial {z}}}}}\right|
_{z=z_{0}+\epsilon }{=\frac{\alpha F_{0}}{gM}}
\end{equation}
These are the same as Eqs.~(\ref{thetatop}), (\ref{thetabottom}), and (\ref
{thetamiddle}) of the main text.

%****************************************************************

\begin{figure}[tbp]
\caption{Motion of the suspension and the test mass under the action of the
generalized force conjugate to the readout variable $x_{{\rm readout}}=x_{%
{\rm testmass}}+\protect\alpha x_{{\rm fiber}}$.}
\label{generalizedforce}
\end{figure}

\begin{figure}[tbp]
\caption{Suspension thermal noise with and without compensation. The dotted
line is the usual uncompensated suspension thermal noise. The dashed line is
for optimal compensation and a sensor that is infinitely precise. The solid
line is for thermal noise compensation that is partially impeded by the
sensor noise. For the case in the figure, the sensor is positioned in the
middle of the suspension fiber; therefore, only violin modes with an even
number of nodes get affected by the compensation. By shifting the sensor
away from the fiber's middle one could compensate thermal noise in the
``odd'' violin peaks as well as the even ones.}
\label{graph}
\end{figure}

\begin{figure}[tbp]
\caption{Optimized value of $\protect\alpha$, for the case when the sensor
is perfect (dashed line) and when the sensor is intrinsically noisy (solid
line); cf.~Eq.~(\ref{alphanew}).}
\label{optimizedalpha}
\end{figure}

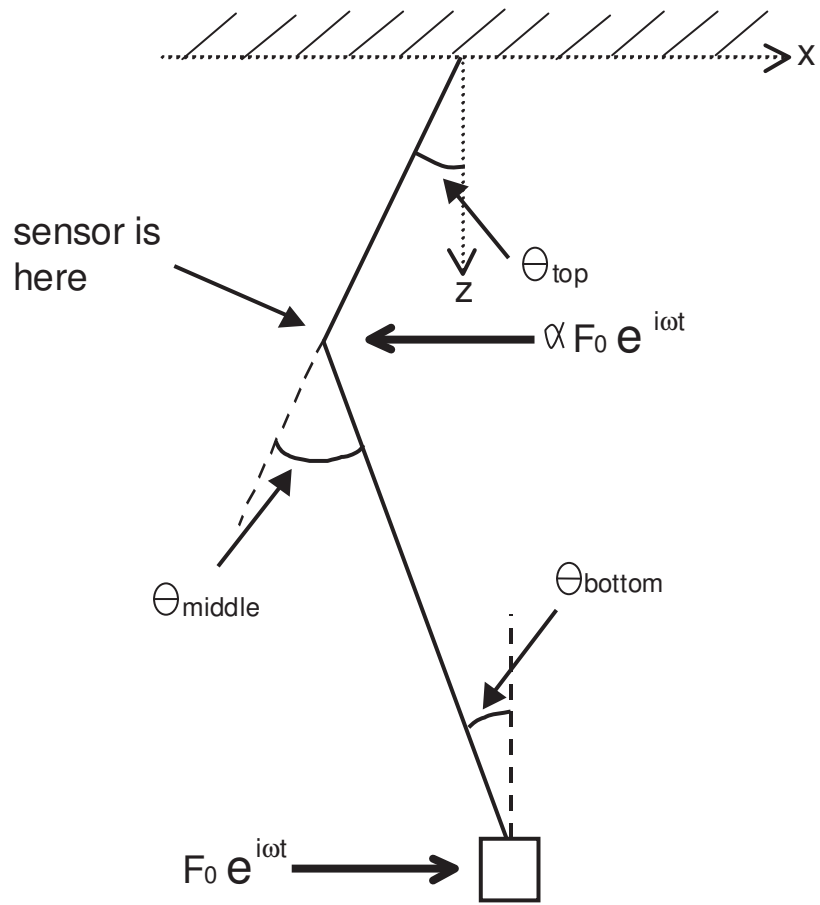
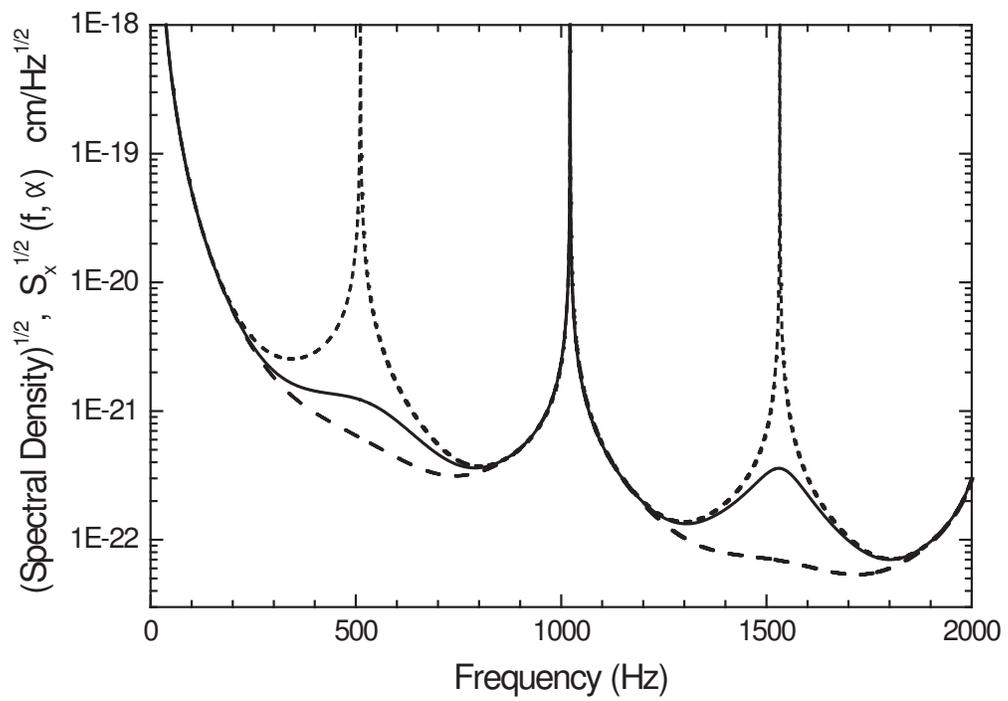
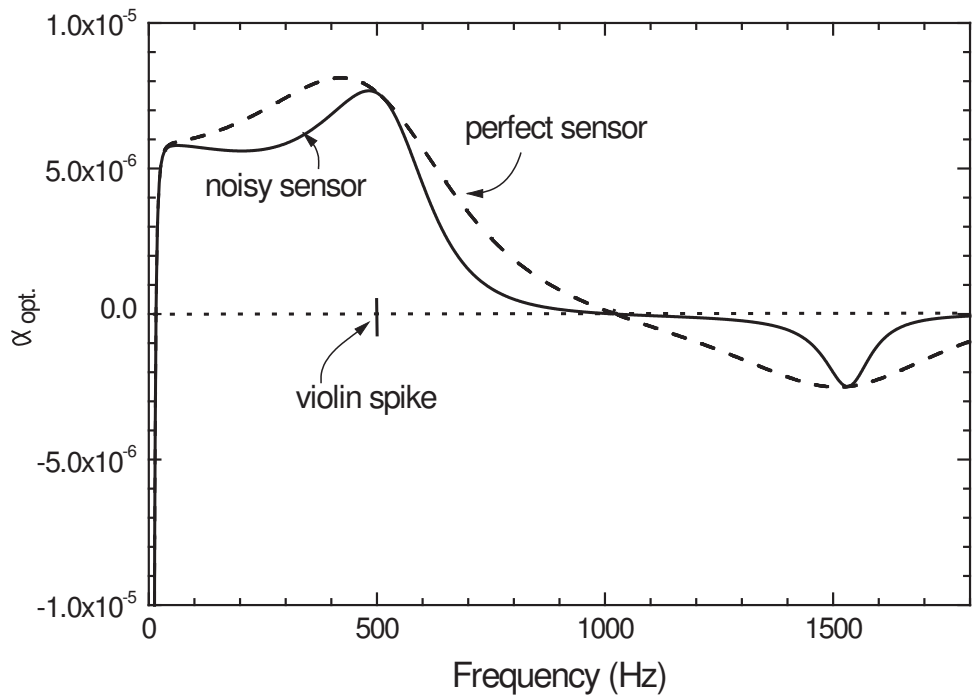
\end{document}